\documentclass[review]{elsarticle}
\usepackage{color}
\usepackage{hyperref}

\begin{document}

\begin{frontmatter}

\title{Imbalance of pairwise efficiency in urban street network}

\author{Minjin Lee}
\address{Department of Energy Science, Sungkyunkwan University, Suwon, Korea}
\author{SangHyun Cheon}
\address{Department of Urban Planning and Design, Hongik University, Seoul, Korea}
\author{Sungmin Lee\corref{mycorrespondingauthor}}
\address{Department of Physics, Korea University, Seoul, Korea}

\cortext[mycorrespondingauthor]{Corresponding author}
\ead{jrpeter@korea.ac.kr}

\begin{abstract}
We investigate how efficient each area of urban street network is connected. Using the detour index, geographic features of street network and pairwise efficiency is studied. To do so, the detour index of 1,832,118 travel route pairs for 85 global cities are explored. We show that the detour index in urban street network is strongly dependent on the angular separation of a pair respective to the city center and that it is a unique property introduced by the intrinsic core-periphery structure of urban street networks. Lastly, a new way of mapping the street network to visualize the overview of efficiency level is proposed.
\end{abstract}

\begin{keyword}
\texttt{Street network, detour index, pairwise efficiency}

\end{keyword}

\end{frontmatter}

\section{Introduction}

 With the continuing urbanization throughout the world, street networks have played a great role to develop urban areas by mediating interactions of people and goods. As the interaction and connectivity becomes denser in an urban area, the efficient road system has been considered an important issue for the sustainable development of a city~\cite{Xu2016}. In order to address the issue, efficient street networks have been actively studied in various research fields. In particular, for last decades they has drawn great interest in physics while the rapid development of complex network provides a powerful tool to analyze macroscopic behavior of complex interactions occurring on the street network \cite{PhysRevE.73.036125,0295-5075-91-1-18003,TRAVENCOLO200889,Barthelemy20111}. A number of studies~\cite{PhysRevE.73.036125,Louf28052013,Levinson2009732,1742-5468-2006-01-P01015,WANG2011285} measure the efficiency level of street networks and give insights for the macro trend of entire network or the distribution of efficiency of each road segment. 

Despite of their great contributions to understand statistical properties of urban road efficiency, previous studies with node based efficiency measure mainly focused on pointing out efficient or inefficient areas. However, it fails to provide the information of which areas are efficiently connected. Such information is practically important for policy making, and deeper understanding on the interactions between subregions is essential to identify which linkages should be improved. In fact, in real world one can easily agree that a node (or a sub region in broad term) cannot be connected to all other sub regions with a same level of efficiency. Even for an area known for high efficiency, it can be efficiently connected to some regions, while it can be also poorly connected to other regions. In general networks, such heterogeneous link weight might not be an important issue, but in street network which should concern its geographical properties the heterogeneity of link weights describes the state of imbalanced efficiency throughout the urban street network. In this paper, we call such phenomena as imbalanced pairwise efficiency in urban street networks and this is our main focus in the present study. 

In order to investigate the pairwise efficiency, we use the detour index \cite{Louf:2013ii} which is the ratio between the travel distance and the geodesic distance (crow-flies distance) of a given origin-destination pair (OD). It is one of the popular and simple ways to measure the efficiency and widely used with various names such as circuit\cite{Levinson2009732}, route factor\cite{morphogenesis}, detour factor\cite{WITLOX2007172,JUSTEN2013146}, straightness \cite{PhysRevE.73.036125,WANG2011285,doi:10.1068/b32045}, route-length efficiency\cite{aldous2010}, and so on \cite{PhysRevE.71.036122}. The issue of detour index is discussed regarding efficiency of travel diary of individuals in transportation studies \cite{WITLOX2007172,JUSTEN2013146,warnes_detour} and it is also used as a proxy to measure the efficiency of road network from the complex network perspective \cite{PhysRevE.73.036125,Levinson2009732,WANG2011285,Louf:2013ii,doi:10.1068/b32045,aldous2010,PhysRevE.71.036122}.  The detour index throughout different cities shows universal behaviors to a certain extent: (1) the average detour indexes across various cities are very comparable \cite{WITLOX2007172} and (2) the detour index tends to decrease as increase of the geodesic distance between OD pairs \cite{Levinson2009732,WITLOX2007172,JUSTEN2013146,Rietveld:1999ky}. The universal pattern must be strongly related to some universal properties of embedded street networks~\cite{Helminen2007331,Rietveld1999}. However comparing the large volume of literature regarding detour index and street network themselves, one rarely find the analysis connecting them to investigate how structure of street network affects its efficiency level \cite{SPADON201718}. 

We aim to understand patterns of pairwise efficiency in urban street networks using detour index of travel routes and discover the relationship between efficiency level and underlying street network structures. We expect to find the fundamental properties of street networks generating heterogeneous efficiency level among subregions. Additionally we suggest a coarse-grained street network which encodes efficiency level among all subregions as an application. The network offers us at-a-glance understanding of the state of imbalanced pairwise efficiency in a city. 

\begin{figure}
\begin{center}
\includegraphics[width=1.\linewidth]{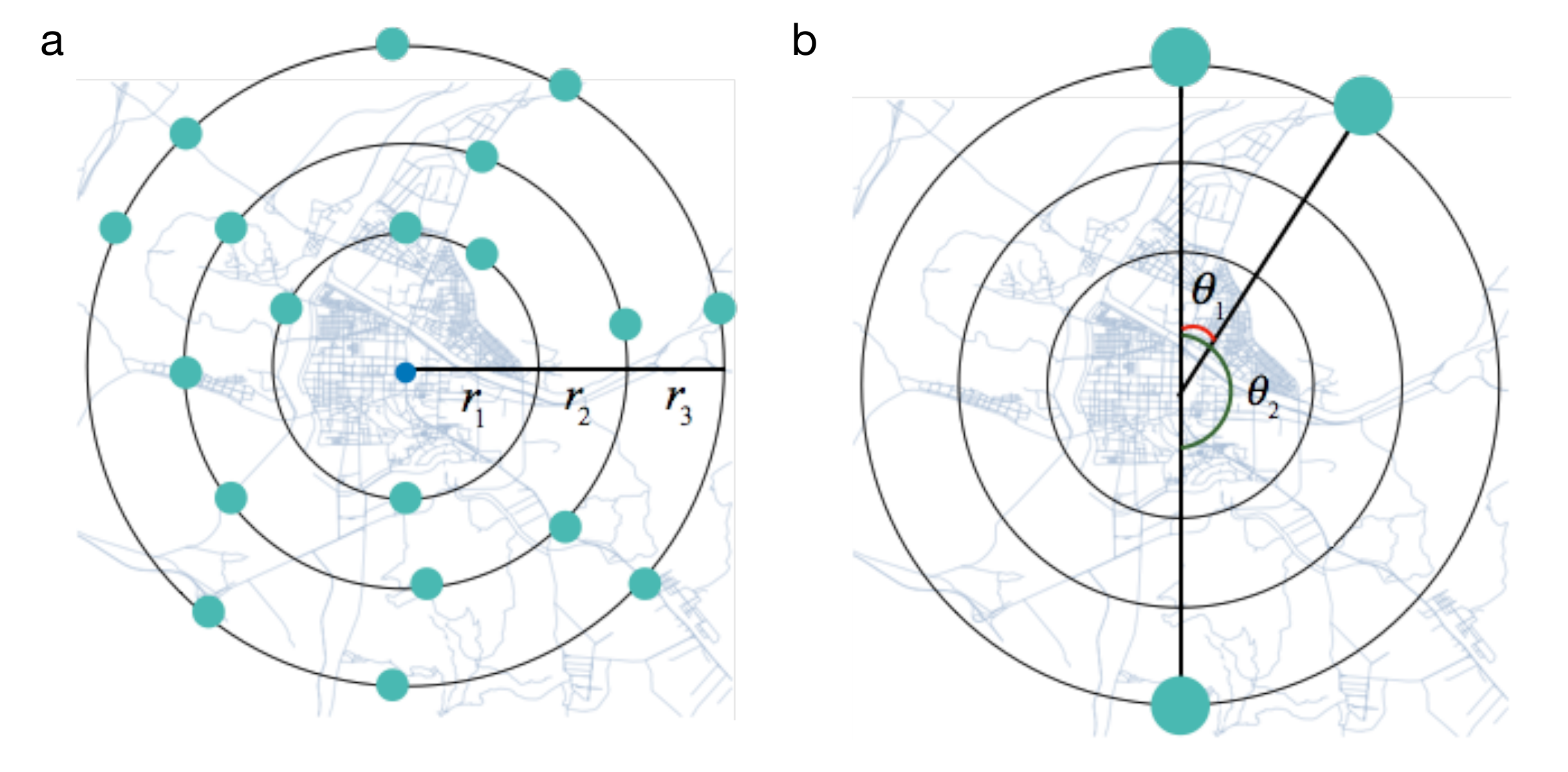}
\end{center}
\caption{(Color online) Sampling OD pairs (a) Points are randomly selected within a circular line of given radius, $r_n$. We make the OD pairs by connecting any two points of same radius. (b) We measure the angular separation,$\theta_n$ between two points and the center point.}
\label{sampling_od}
\end{figure}

\textcolor{blue}{
}

\section{Detour index of empirical route}
\paragraph{Sampling travel routes and measure of detour index}
We analyze 1,832,118 travel routes for 85 global cities. For each city, we collect around 24,000 travel routes (Detail statistics are included in SI) which are shortest routes retrieved using Openstreetmap API ~\cite{OpenStreetMap}. To get more systemic understanding on inhomogeneous efficiency in street networks, we use radius-fixing sampling method ~\cite{Lee2017}. It is selecting origin-destination (OD) pairs on the circular line with having a city center as the center of the circle (Fig.~\ref{sampling_od}a). This sampling enables us to observe the behaviors of travel routes by same radius positions.
In our study, we draw the circular line with various radii, $r_n$, (2$km$, 5$km$, 10$km$, 15$km$, 20$km$, and 30$km$), and generate around 4,000 OD pairs per $r_n$, connecting two points on the circular line in all-to-all.

\begin{figure}
\begin{center}
\includegraphics[width=1.0\linewidth]{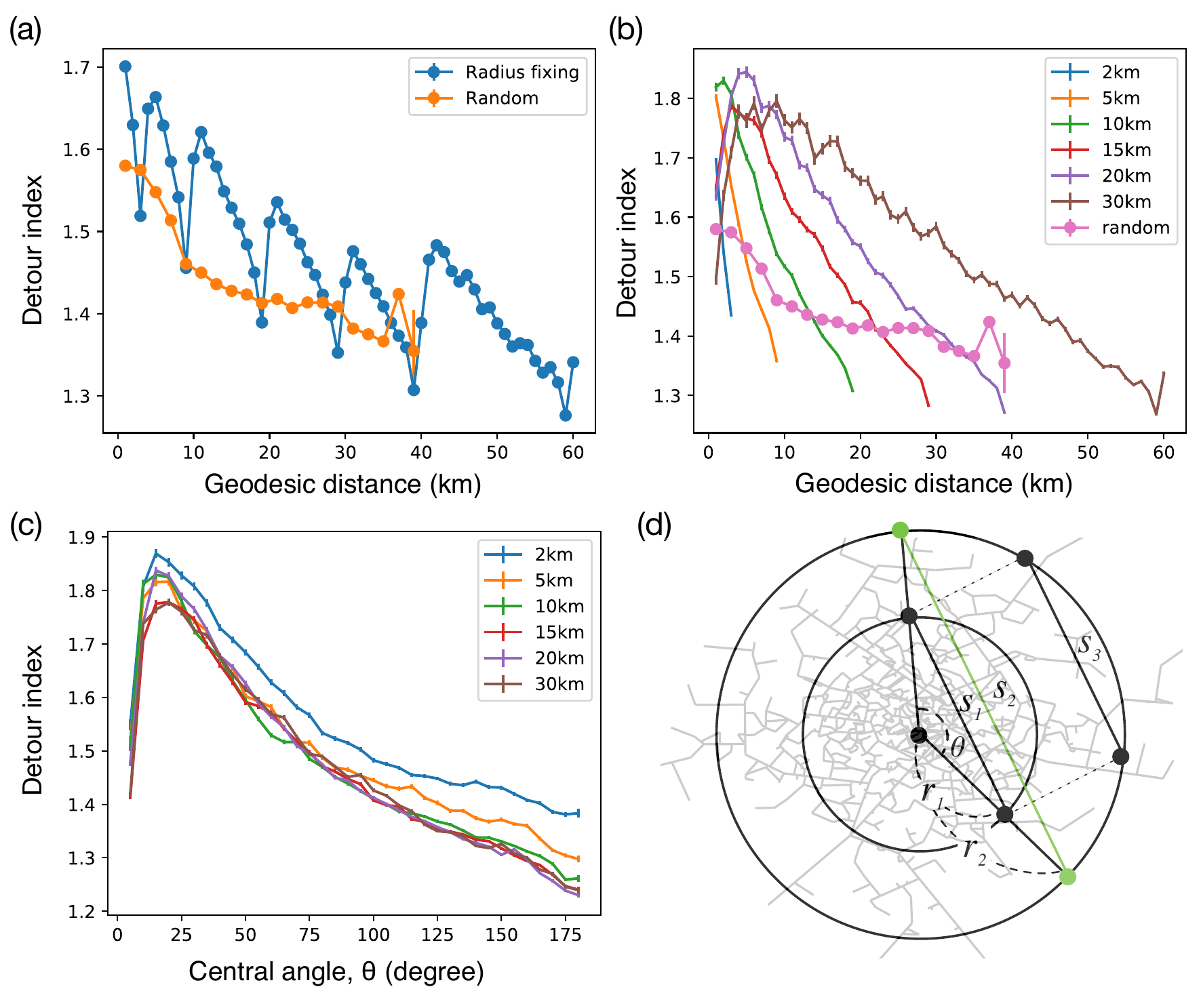}
\end{center}
\caption{(Color online) Average detour index. (\textbf{a}) Average detour index by $s$; (\textbf{b}) Average detour index by radius of route pairs from the center; (\textbf{c}) Average detour index of each radius by $s$; (\textbf{d}) Average street density by radius and average accessibility level by radius.}
\label{average_df}
\end{figure}

Then, we measure the detour index of all sampled routes. Detour index ($DI$), as an metric to quantify the efficiency of street network, is simply ratio between network distance and euclidean distance. 
The detour index of each OD pair from our sample travel routes is 
\[DI_{od}=\frac{d_{od}}{s_{od}}\], where $d_{od}$ is the shortest distance between OD through the street network and $s_{od}$ is the geographically euclidean distance between OD. The higher the $DI$ is, the less efficient the path is. To compare, we additionally select OD pairs which connect any spatially random two points within the boundary of 20$km$ from the center and find the shortest path of them. we call this samples as random samples.

\paragraph{Average trend of detour index}

We first observe the general pattern of average $DI$ in Fig.~\ref{average_df}(a). The $DI$ of random samples continuously decreases as the geodesic distance, $s$, between OD increases (Fig.~\ref{average_df}a Orange line). It is well known trend discussed in previous literature \cite{Levinson2009732,WITLOX2007172,JUSTEN2013146,Rietveld:1999ky}. 
However, the average $DI$ of the radius-fixing samples (Fig.~\ref{average_df}a blue line) shows a periodic peak-and-valley pattern with the decreasing trend, describing $DI$ is not a simple function of $s$. Looking carefully, one can notice the periodicity seems to be related to the radius, as the geodesic distances corresponding to each valley (4$km$, 10$km$, 20$km$, 30$km$, 40$km$, 60$km$) accord with the largest $s$ values which can be possibly generated from each radius (2$km$, 5$km$, 10$km$, 15$km$, 20$km$, 30$km$). 

To elaborate this periodicity, we further plot the same graph by groups of each radius. In Fig.~\ref{average_df}b, each curve represents average $DI$ of routes belonging to each radius group. The first thing to note is the average $DI$ curves of each radius show certainly different values within a same geodesic distance range. Although each radius curve has different $DI$ values, the shape of curves are similar for each other showing log normal like shape. It tells the behavior of the decrease against $s$ exist overall, but the effect of the behavior works differently by the radius of pairs.  The common pattern of the curves shows $DI$ increases for small $s$ and decreases after $s$ gets large. When $s$ is small, $s$ between OD is short enough for many pairs to be directly connected ($DI=1$). The effect of directly connected pairs makes low $DI$ on average in small $s$ range \cite{aldous2010}. The decreasing pattern for large $s$ is due to negative relation between $DI$ and $s$.

We observe the pattern of $DI$ against central angle, $\theta$ in Fig.~\ref{average_df}(c).  Contrary to the curves against $s$ in Fig.~\ref{average_df}(b), the curve of each radius is almost collapsed, although the $DI$ plot for 2km radius is higher than other curves. It describes that the $DI$ value is rather dependent on the angle of a given pair than its geodesic distance in urban street networks. The dependency is strong enough that pairs in similar angular position show similar efficiency level regardless of their locations or distance to the center. In practice, $\theta$ can be understood as a connecting direction of a pair, whether it is the connection through the core or the connection between peripheries. As the larger $\theta$ is, the more connecting direction is toward the center (See an example in Fig.~\ref{average_df}(d)). The pairs $s_1$ and $s_2$ are with a same angle ($\theta$) and $s_1$ and $s_3$ are with a same geodesic distance. While both $s_1$ and $s_2$ are located to naturally pass through the core, $s_3$ with smaller angle than $\theta$ is located to connect peripheries.  The fact that $DI$ is negatively related to the $\theta$ ($>20^{\circ}$) tells travel routes are more efficient when they have to pass the city center rather than travel between peripheries. It is noteworthy that the dependency on the central angle is shown as the dominant behavior throughout most individual city (Supplementary Figure 1 and 2).



\paragraph{Core-periphery property of street networks}

\begin{figure}
\begin{center}
\includegraphics[width=1.0\linewidth]{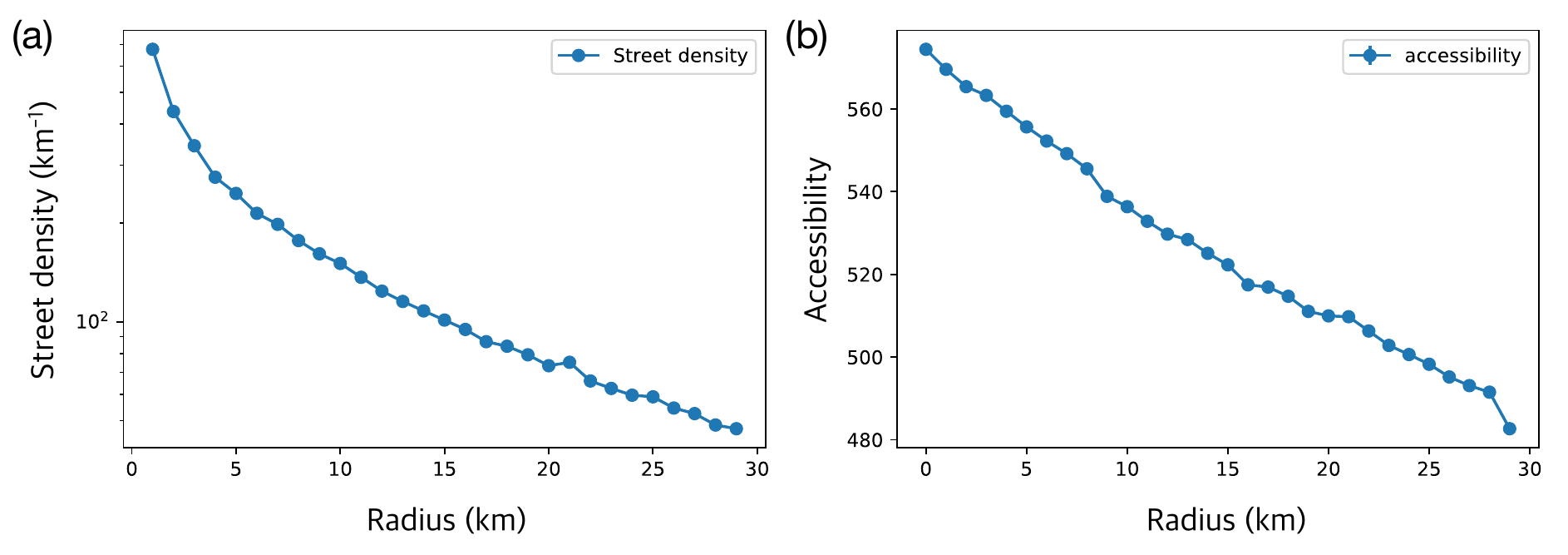}
\end{center}
\caption{(Color online) Average pattern of street structure. (\textbf{a}) Average street density by $r$ (semi log plot); (\textbf{b}) Average accessibility by $r$}
\label{core-periphery}
\end{figure}

From the previous observations, we know that the efficiency level is determined depending on the connecting direction of a pair. It suggests that a general urban road structure is designed to have good connections toward the center and poor connections between peripheries. Related to this, \cite{Lee2017} suggested that many urban street networks have a geographical core-periphery structure which let travel routes bended toward the center. We verify if the core-periphery structure is the key factor to generate the systemic inhomogeneous efficiency by analyzing street network structure and artificial street models. In fact, the geographic, functional or economic core-periphery urban structure has been studied in a number of literature \cite{Pain2008_coreperi,christaller1966central,losch1954economics,spatialeconomy} for decades. \cite{Clark:2016vj} proposed an empirical evidence of monocentricity of urban areas by showing exponentially negative relationship between distance from the center and population density and a few studies discussed core-periphery structure from the complex network perspective \cite{PhysRevE.89.032810,PhysRevE.72.046111, Rothrsif20120259}. However it is neither simple nor intuitive to apply the existing core-periphery methods developed for complex network analysis because of the unique properties of street networks such as relatively homogeneous degree distribution. Thus, as the alternative to measure core-peripheriness, we pay attention to the variation of street density and accessibility on street networks by distance to the center. 

We consider that the areas of high density of streets are likely to be cores \cite{PhysRevE.72.046111}. It is based on the literature detecting core-periphery using random walkers \cite{Rossa:2013do}. When street network is geographically coarse-grained into the network of subregions (which will be discussed in section 4), a sub region with many intersections inside would have high edge weight, and it will increase the chance of random walkers to visit \cite{Noh2004,Lee2009}. According to the \cite{Rossa:2013do}, subregions with high density of roads could be cores. To quantitively validate if the urban street structure does have a geographical core-periphery pattern, we plot the street density gradation by radius averaged for the 85 cities' street networks in Fig.~\ref{core-periphery}(a). It shows exponential decaying behavior. 
Next, the accessibility, which is less intuitive than the density, usually measures how well street segments is connected to each other. We suppose that a high accessibility area is likely to be a core. Based on the idea of core-periphery in complex networks \cite{Rombach:2017bh}, core nodes should be well connected with both other core nodes and periphery nodes but connections between periphery nodes should be poor. Because core nodes are likely connected to more various levels of nodes than periphery nodes, they would have higher accessibility level than periphery nodes. Among various ways to measure accessibility of a network, \cite{TRAVENCOLO200889} and  \cite{1742-5468-2006-01-P01015} especially define accessibility of a node, $A$, by measuring diversity of reached node from self-avoiding random walk on street networks. 
Using the self-avoiding random walks on the street networks of our sampled cities, we identify the destination node, $j$, after 20 steps from random starting point, $i$. Then we define the probability that node $j$ is reached from node $i$ as $P_h(j,i)$ and calculate the Shannon entropy of the final destinations, $j$ like below. We conduct 1000 random walk simulations and average the values for every node $i$ in the networks. We group $A$ for each radius range and average them for all 85 cities (See Fig.~\ref{core-periphery}(b)).
\[A(i)=exp\bigg(-\sum_{j=1}^{N} P_h(j,i)\log(P_h(j,i)\bigg)\]
As shown in Fig.~\ref{core-periphery} (b), the accessibility is linearly decreasing by radius. Streets are generally denser and the possibility to approach various parts gets higher if it is close to the geographical center of city. It proves urban street networks on average have the core-periphery structure. Same observation for individual city also shows that most of cities have a core-periphery structure (Supplementary Figure 3 and 4). 

\begin{figure}
\begin{center}
	\includegraphics[width=1\linewidth]{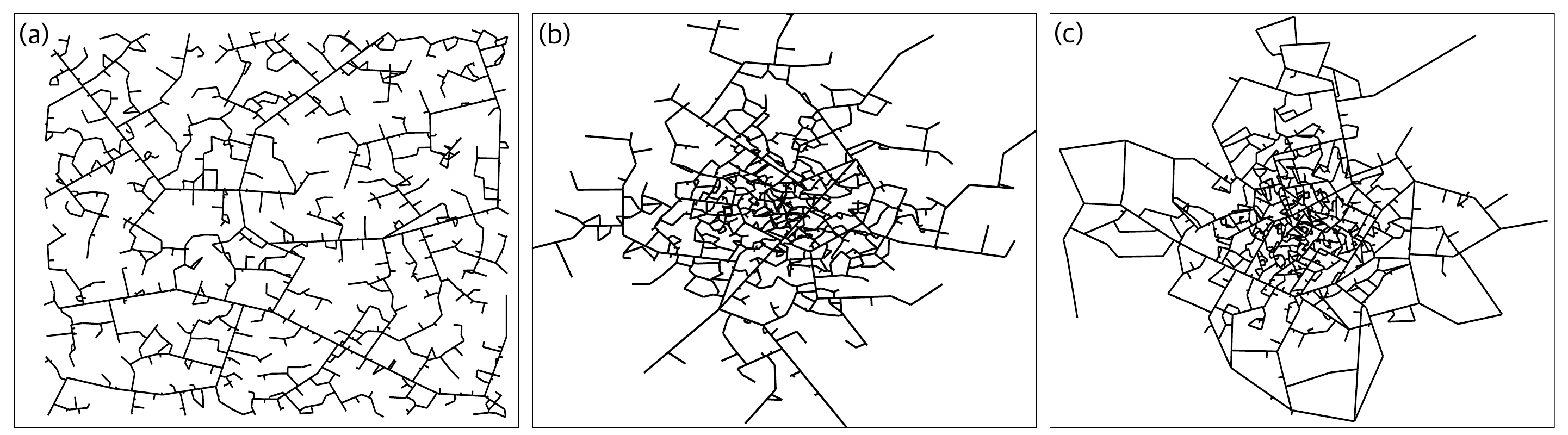}
\end{center}
\caption{(Color online) Street models with randomly distributed intersections (a), exponentially distributed intersections (b,c). Road threshold is much larger in (c) than (b).}
\label{density_street_model}
\end{figure}

\section{Street network model}
\paragraph{Building street model}
Based on our analysis using empirical data, we build the hypothesis that core-periphery structure of urban street networks generates the strong angular dependency of $DI$. In order to validate the hypothesis, we build an artificial street model. We modify the street evolving model suggested by \cite{PhysRevLett.100.138702}. Basically, a street network evolves by adding street intersections and connecting the intersections. From initially distributed intersections which is geographically random, we connect them at random to make an initial spatial graph. Here, overlapped edges are not allowed. For each time step, we add $n$ street intersections of which positions are selected from a given density distribution function $P(r)$. Added intersections are connected to the closest point of existing network, as long as the new edge is not longer than a threshold, $h$, we set. If more than two intersections are connected to same points, a new bridge node is introduced in-between so that they are connected with shortest length to reduce the construction cost. Once new nodes are connected to existing networks during several iterations, we connect the end nodes of the network if the newly made edge is no longer than the threshold $h$. The process is to make loops in the network, avoiding unrealistic tree-like networks. Our generated networks have similar statistical properties with real street networks in terms of road length distribution, area size distribution, and phi distribution which \cite{LAMMER200689} mentioned as intrinsic urban street network properties. See Supplementary Figure 5. 

We validate the hypothesis by varying density distributions and threshold values. We first use exponential-decaying function of the intersection density distribution, in which the positions of intersections follow $P(r) = \exp(\frac{-|r|}{r_{0}})$. Here, $r$ is the distance from the center and $r_{0}$ is the mean distance. As a null model, we also use randomly uniformly distributed intersection positions within a boundary. We vary the threshold values as $h=4$, 6, 8, 10, 12, 60 for the box boundary of 60 unit. By tuning $h$ we test the change of $DI$. Fig.~\ref{density_street_model} shows the examples of street networks under different conditions; a: intersections randomly distributed (RD), b: exponentially distributed intersections (ED), c: exponentially distributed intersection and higher road segment threshold. As shown in the examples, the street networks look different depending on both the density distribution and the threshold. We present their statistics and detail analysis in the following section.

\paragraph{Model result}

\begin{figure}
\begin{center}
\includegraphics[width=1.0\linewidth]{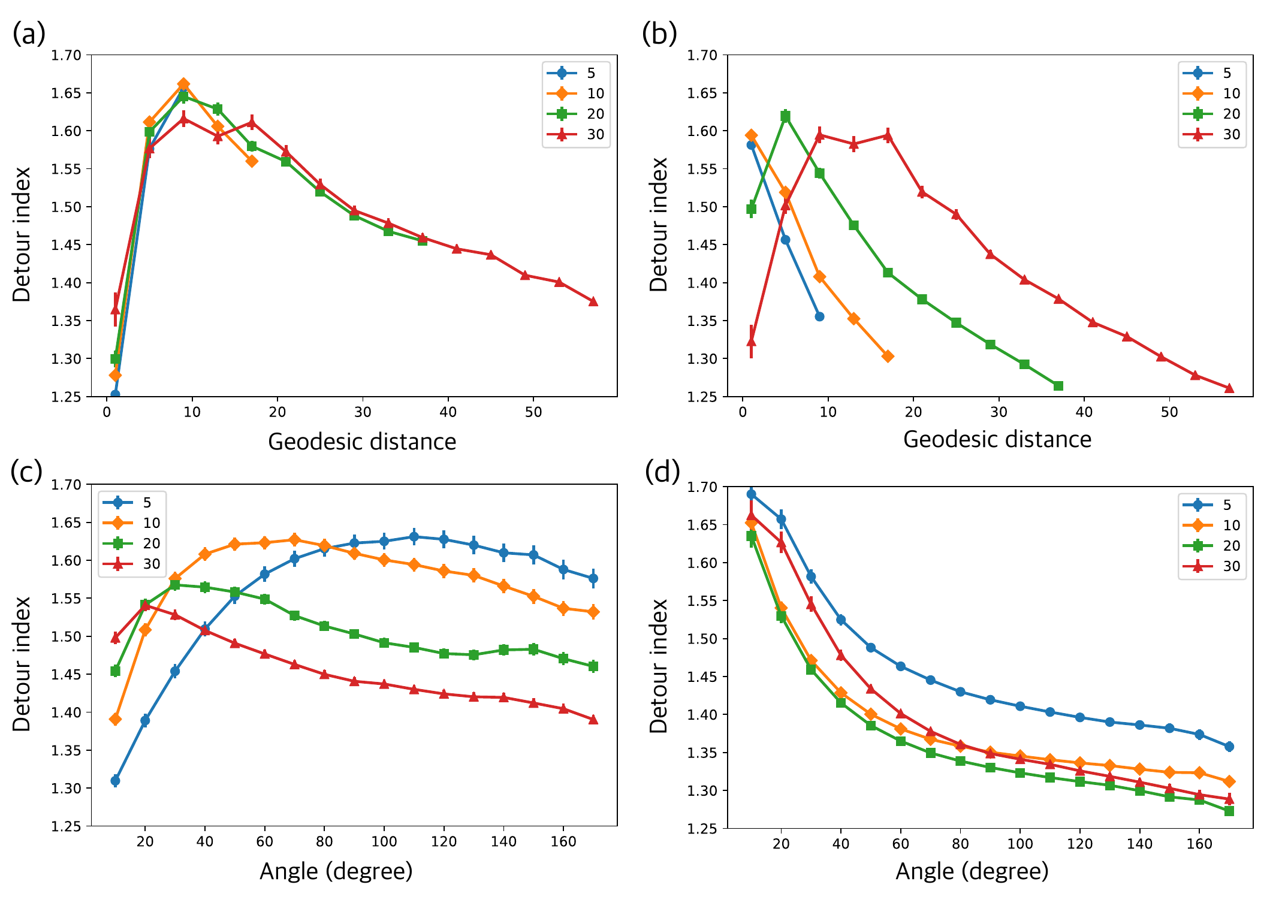}
\end{center}
\caption{(Color online) Model result. Average detour index of several radii by $s$ for RD (a) and ED (b). Average detour index of several radii by $\theta$ for RD (c) and ED (d). Used radii are 5, 10, 20, and 30 unit.}
\label{model_result}
\end{figure}

On the generated street networks, we collect shortest paths of selected OD pairs using radius-fixing sampling method. Then, we look into the average detour index for geodesic distance $s$ and for central angle $\theta$.
Fig.~\ref{model_result}a shows the average trend of $DI$ by $s$. While $DI$ generally decreases as $s$, the periodic peak-and-valley pattern which we observed in empirical networks is only shown in the ED networks, not in our null model (RD). It tells the radius fixing sampling does capture the unusual behavior of empirical networks rather than display mere artifact from sampling method. We, then, also segregate the data into each radius group to see its dependency on radii. Curves of all radii are collapsed in RD networks, describing there is no radius dependency but only $s$ dominantly determines $DI$ (Fig.~\ref{model_result}a). On the contrary, the curves in ED networks are not overlapped and various $DI$ values are shown for a same $s$, (Fig.~\ref{model_result}b) as the empirical networks do. Next, we observe the relation between $DI$ and $\theta$. Curves of each radius in RD networks are different each other (Fig.~\ref{model_result}c), while plots in ED networks show similar shapes and mostly collapsed (Fig.~\ref{model_result}d). The model result also tells dependency of detour index on the central angle is stronger for ED networks. 

By comparison with the null model, it becomes evident that core-periphery structure makes detour index more dependent on the angular position than the absolute geographical distance in urban street networks. As seen in result of RD network (Fig.\ref{model_result}(a)), when roads are equally and uniformly distributed, only the $s$ affects the efficiency of the route. However in real networks where it is not uniform, the pattern of road networks works stronger than the $s$. In particular, our experiments reveal that any routes passing through the center. In other words, routes with higher central angle are far more efficient in core-periphery structure. 

\begin{figure}
\begin{center}
\includegraphics[width=0.6\linewidth]{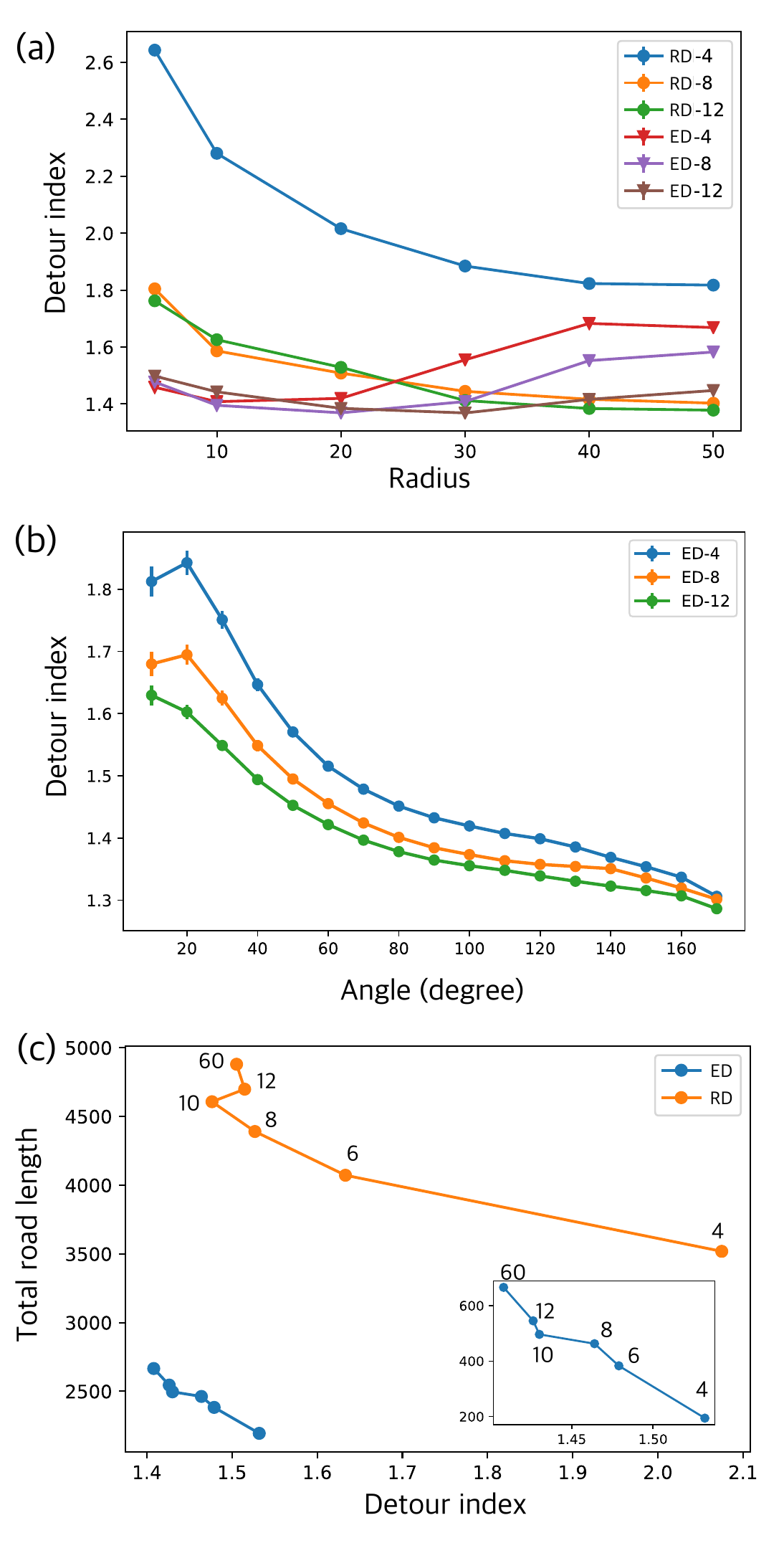}
\end{center}
\caption{(Color online) Improvement of efficiency by changing the threshold $h$. (a) Average detour index of both RD and ED by radius for various threshold (h=4,8,12). (b) Average detour index of ED by angles for various threshold (h=4,8,12). (c) Total road length vs detour index. The inset is zoomed plot of ED case.}
\label{model_result_acc}
\end{figure}

Hence, how can we relieve the imbalance of street networks and increase the efficiency? Improving street network is concerned with increasing efficiency and accessibility with minimum cost of construction. We generate several street networks with different threshold $h$ of road length to see how increase of the threshold affects the detour index. Fig.~\ref{model_result_acc}a illustrates average detour index against radii and their changes with various $h$. As seen in our previous result, the curve of RD decreases by the radius because $DI$ and $s$ are negatively related. However in ED networks, although $s$ of large radius tends to be large and should make detour index lower, $DI$ rather gets higher or does not change in large radius. It means areas located in large radius are not well connected. This phenomena is clearer with low threshold, $h$. With low $h$, streets are rarely built between intersections in peripheries as the distances are mostly beyond the threshold distances. This imbalance is relieved by increasing $h$. Detour index in large radii gets smaller by increasing $h$. Fig.~\ref{model_result_acc}(b) shows that $DI$ particularly decreases for small angular routes. It clearly describes that increase of $h$ improve the connection between peripheries located far from the center. In RD networks, we also find the improvement of efficiency with increase of $h$. However it does not change the shape of curves, but only shift the curve downward. It tells varying threshold length affects differently by network type. Contrary to the ED networks where extending $h$ allows intersections in peripheries more connected, increase of $h$ in RD networks just makes more roads between all intersections.

Although we expect increasing threshold length improves efficiency to a certain extend, in real world the problem is a cost for the additional road length. Fig.~\ref{model_result_acc}c shows how detour index and total road length in average changes by the threshold (threshold length is the number written in each point). As one can expect, increase of threshold length mostly causes longer total road length and lower detour index. Notably, for RD networks, there is a optimal threshold ,10, more than in which efficiency does not get better and total road length also does not significantly increase. ED networks does not seem to have obvious optimal point, but after threshold 10 the slope of the curve slightly get steeper. We can understand the effect of expending road length becomes weaker from threshold 10 unit.  
 
 \section{Applications: Local and pairwise efficiency network}
 To expand the study of imbalance of efficiency level, we generalize our sampling method and measure the pairwise efficiency among all subregions in a city. We build a spatial efficiency network which is based on coarse-grained street network. As a case study, we analyze 20$km$ x 20$km$ square size of Seoul (We additionally analyze several big cities such as Houston, London, Los Angeles, Phoenix, Paris, Mexico City, Shanghai, Sao Paulo, Washington D.C. (Supplementary Figures 6-12)).
 
\begin{figure}
\begin{center}
\includegraphics[width=1\linewidth]{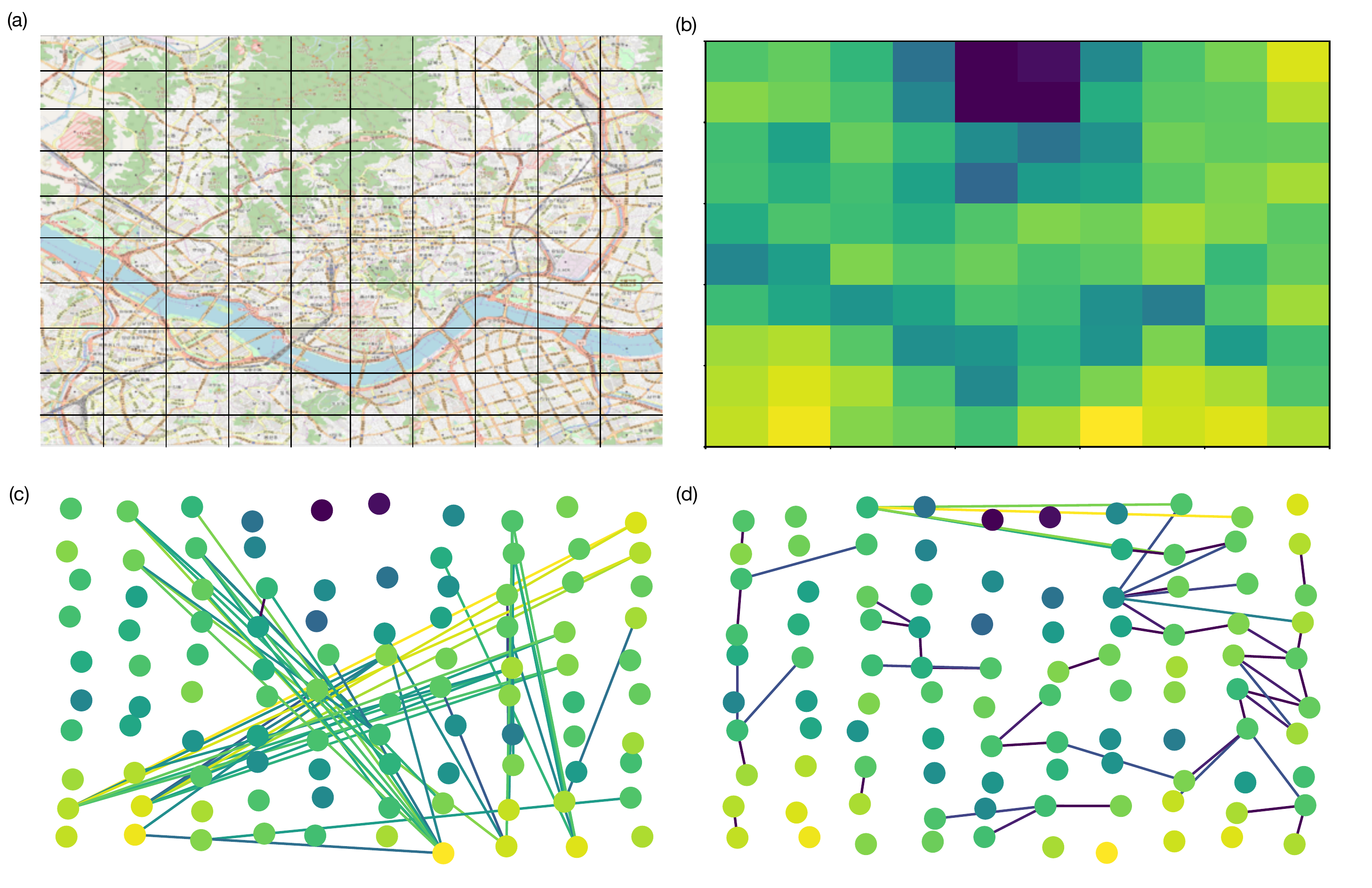}
\end{center}
\caption{(Color online) Local efficiency network in Seoul. (a)Boundary of the sampled map of Seoul and the division into 10x10 cells. (b)Node strength in Seoul. Each cell represent the boundary that each node covers. We sum the efficiency level of edges corresponding to each node and assign it as the strength. The brighter the cell is, the more efficient the area is. (c)Sub-network of the selected 100 most efficient edges. Brighter edge colors represent relatively higher efficiency (d)Sub-network of the selected 100 worst efficient edges.} 
\label{local_efficiency_map}
\end{figure}

 We divide the city space into 10 by 10 grid cells as shown in Figure.~\ref{local_efficiency_map}(a) and build an all-to-all network whose node is centroid point of each cell and edge weight is an efficiency level between corresponding subregions. To know the efficiency level between subregions, we randomly sample the 50,000 shortest routes within the boundary and measure the detour index of sample routes. The routes are grouped for each edge whose node cells include the origin and destination of routes and inversed detour indexes in a group are averaged and assigned as the corresponding edge weight.

Firstly, Fig.~\ref{local_efficiency_map}(b) shows the node strength, which is the sum of edge weights of nodes, the subregions. Higher node strength is colored brighter and represents the node is connected to other place efficiently. By the nature of $DI$ on urban street networks, when a pair has longer geodesic distance and larger central angle, $DI$ would be smaller. The visualization in Fig.~\ref{local_efficiency_map}(b) captures such behavior well throughout Seoul. As the most efficient areas, two bright agglomerations on bottom (Gangnam and Yeongdeungpo area) are shown. Since they are located in the corners, which let them have longer geodesic distance pairs and lower $DI$, it seems natural for them to be considered as efficient areas. However it is not only a result of systemic artifact of $DI$ metric. If Seoul street structure were exactly isotropic, four corners of the boundary would be equally the highest strength areas. The disparity in the strength between the four corners rather exhibits the intrinsic characteristic of Seoul street structure influenced by policy, geographical constraints, and so on. In fact, the Gangnam area is known as a new subcenter of Seoul receiving a great benefit of infrastructure investment and Yeongdeungpo is also likely to enjoy the benefit of good infrastructure since it is next to the Yeoyido which is other subcenter of Seoul. \cite{KIM2012142}. The outstanding darker areas on middle-top and across middle of the city are natural barriers, Bukhan mountain and Han river respectively. 

Secondly, we look into the pairwise efficiency. Especially we select 100 efficient pairs and 100 inefficient pairs in Figure~\ref{local_efficiency_map}(c) and (d). The edge color also represents the efficient level that brighter color means higher efficiency. The high efficient edges are mainly connection through the center which is in line with the observation from the average pattern. Interestingly, one can notice that Gangnam is efficiently connected to the areas in various directions such as left-top, core, and even right-top areas. It reveals Gangnam is a particular area efficiently accessible from various areas in Seoul. It makes sense with the history of land development in Seoul. Seoul government had focused on investing road infrastructure in Gangnam area through various policies such as the land readjustment project to disperse the population overcrowded in Northern part of Seoul and build a new subcenter since the 80's \cite{KIM2012142}. Such governmental involvement developed the accessibility toward and inside Gangnam area and generated special topological characteristic in Seoul road network. The 100 least efficient edges in Fig.~\ref{local_efficiency_map}(d) are mostly short distance pairs as we can expect. We omit the pairs placed in the Bukhan mountain and Han river area to exclude the impact of natural barrier. The selected low efficient connections are generally short routes and spread over all areas.

\begin{figure}
\begin{center}
\includegraphics[width=1\linewidth]{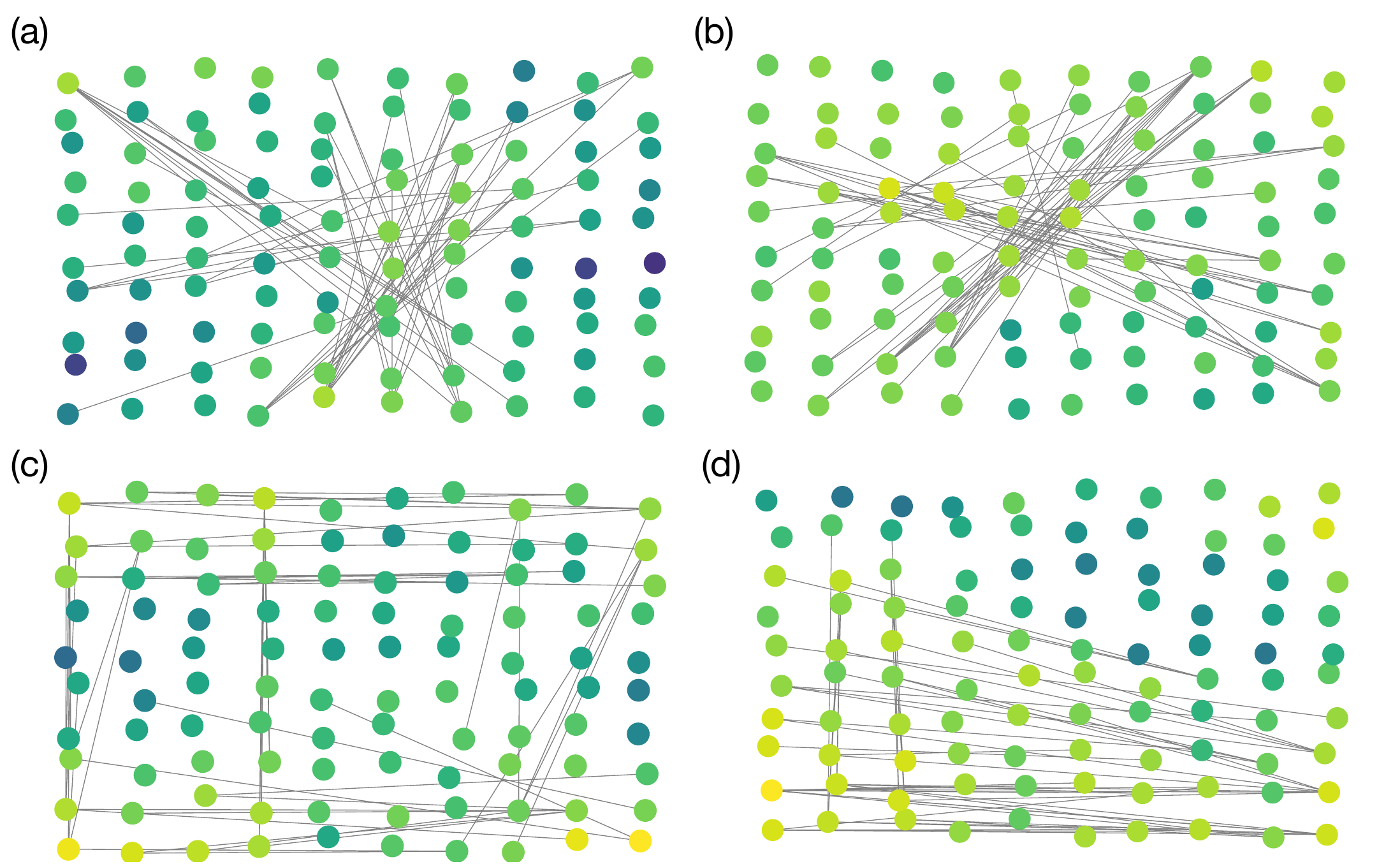}
\end{center}
\caption{(Color online) Subnetworks of 100 high efficiency edges in (a)London, (b)Paris, (c)Houston, (d)Los Angeles} 
\label{local_efficiency_map_4}
\end{figure}

We lastly compare the subnetworks of 100 high efficient edges in four global cities, London, Paris, Houston, and Los Angeles in Fig.~\ref{local_efficiency_map_4}(a)-(d). London and Paris are representative metropolitan areas in Europe which has been evolved from the cores for thousands years, despite of their current polycentric properties \cite{doi:10.1080/02723638.2014.939538}. Houston and Los Angeles are also representative American cities and they are known as planned cities without distinct cores \cite{doi:10.1080/02723638.2016.1200279,GLAESER20042481}, as we expect from the the observed average pattern and monocentric cities, while the American cities do not show such patterns. In the network of Houston, the selected pairs follow along the expressways and Los Angeles exhibits efficient connections are spread out only in half region of the sampled area without any special pattern. It demonstrates that the cities do not have strong distinct cores in the middle but rather a dispersed street structure. The efficiency network is simple but enables us to abstract and visualize the main structural properties of urban street networks. 

\section{Discussion}
We observe the heterogeneous pairwise efficiency patterns in urban street networks using the detour index metric. Changing the sampling method let us discover the intrinsic imbalance existing in the networks. Interestingly, the level of detour index is dependent on the relative angular position of route to the center rather than its geodesic distance or radius from the center.  In other words, travel routes with larger center angles which passing through city center are significantly more efficient than routes with smaller center angles which connecting peripheries.  This trend shown in average data is also found in many cities, implying that urban street networks share a universal structural characteristic and it generates intrinsic imbalance of efficiency level.

We find the heterogeneous pairwise efficiency and the strong dependency on the angular position are related to the core-periphery property of urban street networks. We first quantitatively prove that most of urban street networks have core-periphery structure using distribution of street density and accessibility. Then our model demonstrates the angular dependency is result from the core-periphery structure of networks. To alleviate the imbalance of efficiency, we gradually connect the peripheries by increasing threshold of road segments. increase of threshold generates outer beltways in a city bypassing the busy downtowns and increases the efficiency between peripheries. The model also suggest there seems an optimal threshold minimizing additional marginal road length, implying existence of a possible optimal moment or location to build beltways to maximize the social benefit (global efficiency) against the cost. 

To extend our discussion of pairwise efficiency, we suggest the efficiency network by encoding the $DI$ information on coarse-grained street network. Through the case study of Seoul, the efficiency network reveals both general trend shown in average pattern of $DI$ and unique structural characteristic of Seoul. Especially, it pinpoints Gangnam as the efficient areas with good connections with various other regions. Gangnam is well known sub center in Seoul which has been developed with great investment of road infrastructure. It is worth noting that this simple method captures the flow of efficiency and find a transport center only from street networks. Thus, we expect the efficiency network can be a good tool to uncover complex interactions occuring on the urban street structures and deepen our understanding of urban street network as a footprint of mobilities and economic and social activities in a city. 
 
 By suggesting the notion of pairwise efficiency in street networks and linking the efficiency issue with the structural characteristics of street networks, we provide a new perspective to deepen our understanding of urban street networks. Extending our study into individual city level and further analysis with street hierarchy can reveal the relation between efficiency level and street structure more explicit. 

\section*{Acknowledgments}
SL is supported by Basic Science Research Program
through the National Research Foundation of Korea (NRF) funded by the Ministry of Education
(2016R1A6A3A11932833). ML and SC are supported by Basic Science Research Program through the National Research Foundation of Korea (NRF) funded by the Ministry of Science, ICT \& Future Planning(2016R1A2B4013843)

\bibliography{mybibfile}

\end{document}